\documentclass{iaus}
\usepackage{graphicx}
\usepackage{epsfig}

\title[short title of paper] %% give here short title %%
{A New 3D Potential-Density Basis Set}

\author[short author list]   %% give here short author list %%
{Alireza Rahmati$^1$, Mir Abbas Jalali$^2$}

\affiliation{$^1$Physics Department, Sharif University of
Technology, Tehran, Iran \break email: alistar@sharif.edu\\[\affilskip]
$^2$Department of Mechanical Engineering, Sharif University of
Technology, Tehran, Iran \break email: mjalali@sharif.edu}
 
\pubyear{2007}
\volume{245}  %% insert here IAU Symposium No.
\date{?? and in revised form ??}
\jname{Proceedings Title IAU Symposium}
\editors{A.C. Editor, B.D. Editor \& C.E. Editor, eds.}

\begin{document}

\maketitle

\begin{abstract}
A set of bi-orthogonal potential-density basis functions is
introduced to model the density and its associated gravitational 
field of three dimensional stellar systems. Radial components 
of our basis functions are weighted integral forms of spherical 
Bessel functions. We discuss on the properties of our basis functions 
and demonstrate their shapes for the latitudinal Fourier number $l=2$.  
 

\end{abstract}

\firstsection % if your document starts with a section,

\section{Introduction}
Potential--density basis sets are used for expanding the solution 
of Poisson's equation during N-body simulations and stability analysis 
of stellar systems. Spherical harmonics are the most efficient functions 
for expansions in the azimuthal and latitudinal directions but finding 
bi-orthogonal components in the radial direction, which are expressible 
in terms of ordinary or special functions, is not a straightforward task. 
A numerical method has been introduced by Weinberg (1999) for constructing 
the basis functions in the radial direction. His method always gives a 
bi-orthogonal set whose components oscillate in the radial direction. 
However, the amplitude of oscillations and the locations of zeros of
such functions do not necessarily assure a good performance. Especially, 
when one of the peaks of radial components is higher than others 
[this happens for the Clutton-Brock (1973) and Hernquist \& Ostriker (1992) 
functions], computations are rapidly biased. Here we use Clutton-Brock's 
(1972) idea and construct a basis set using the weighted integral form 
of spherical Bessel functions. The functions that we find have shown a 
good performance in the stability analysis of spherical systems
(Jalali \& Hunter 2007). 

\vspace{-0.15in}

\section{A new potential--density pair}\label{sec:basis}

The Laplace operator is Hermitian, and therefore, its eigenfunctions 
always form a complete and bi-orthogonal set (\cite[Arfken 1985]{Arfken}).
A three dimensional generalization of Clutton-Brock's (1972) method 
suggests to introduce
\begin{eqnarray}
\Phi_{lmn}(r,\theta,\phi) &=& -Y_{lm}(\theta,\phi) \Phi_{ln}(r)=
-Y_{lm}(\theta,\phi) \int_{0}^{\infty}j_l(kr/a)g_{ln}(k)dk, 
\label{eq:potential} \\
\rho_{lmn}(r,\theta,\phi) &=& 
\frac {Y_{lm}(\theta,\phi)}{4\pi Ga^2} \rho_{ln}(r)=
\frac {Y_{lm}(\theta,\phi)}{4\pi Ga^2} 
\int_{0}^{\infty}j_l(kr/a)g_{ln}(k)k^2dk, \label{eq:density}
\end{eqnarray}
as the potential and density basis functions. Here $g_{nl}(k)$ are 
some orthogonal functions and $a$ is a length scale. $j_l(r)$ and 
$Y_{lm}(\theta,\phi)$ are the spherical Bessel functions and spherical 
harmonics, respectively. Their combination $j_l(r)Y_{lm}(\theta,\phi)$ 
satisfies Poisson's equation. It is straightforward to show that 
$\nabla^2 \Phi_{lmn}=4\pi G\rho_{lmn}$. Denoting $L^{2l}_{n}$ as 
Laugerre functions and considering their orthogonality, the choice 
of $g_{ln}(k)=k^l L^{2l}_{n}(2k)e^{-k}$ leads to the following 
bi-orthogonality relation between $\Phi_{lmn}$ and $\rho_{lmn}$:  
\begin{eqnarray}
\int_{-\pi}^{\pi}\int_{0}^{\pi}\int_{0}^{\infty}
\rho_{lmn}\Phi^{*}_{lmn} r^2 dr d\theta d\phi=
-\delta_{l,l'}\delta_{m,m'}\delta_{n,n'}
\frac{a(2l+n)!}{2^{2l+4}G n!}. \label{eq:orthogonality}
\end{eqnarray}
The above potential and density functions oscillate in the 
$r$-direction and the amplitude of oscillations decays slowly. 
Peaks of these basis functions accumulate near the center 
by increasing $n$, and therefore, they are suitable for stability 
analyses that utilize matrix methods. The first potential function 
in the radial direction is $\Phi_{00}(r)=-r^{-1}\arctan{r}$
whose profile falls off similar to $r^{-1}$ as $r\rightarrow \infty$.
Our basis functions can be computed both numerically and analytically. 
Since the spherical Bessel functions and their combinations with 
Laugerre functions decay rapidly, definite integrals in (2.1) and 
(2.2) converge very fast by adopting standard numerical schemes. 
For instance, we have computed and displayed in Figure 1 the radial 
potential and density basis functions for $l=2$ and several values 
of $n$.
\begin{figure}[h] 
  \centering
  \begin{tabular}{cc}
    \includegraphics[width=2.5in]{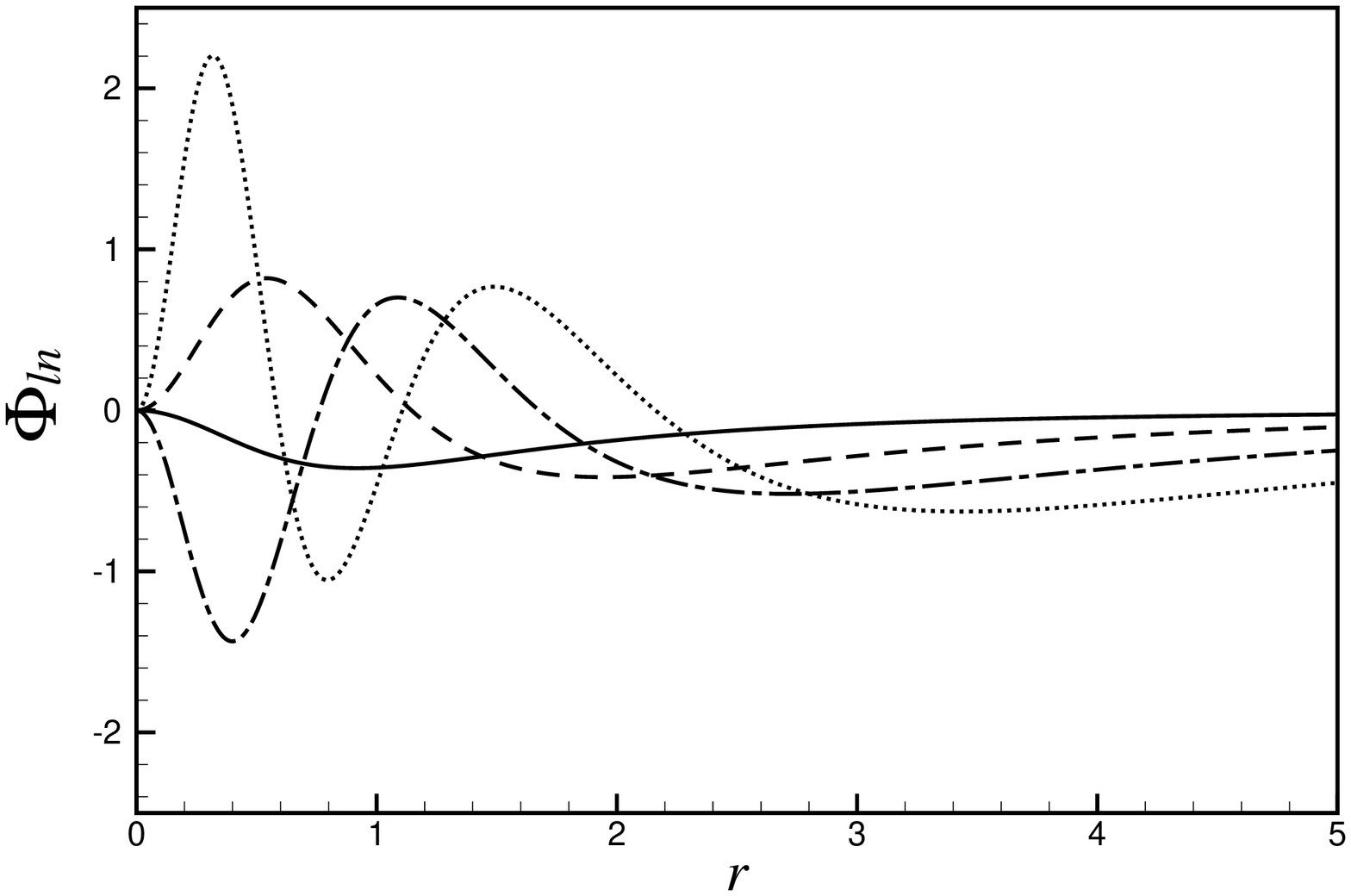}
    \includegraphics[width=2.5in]{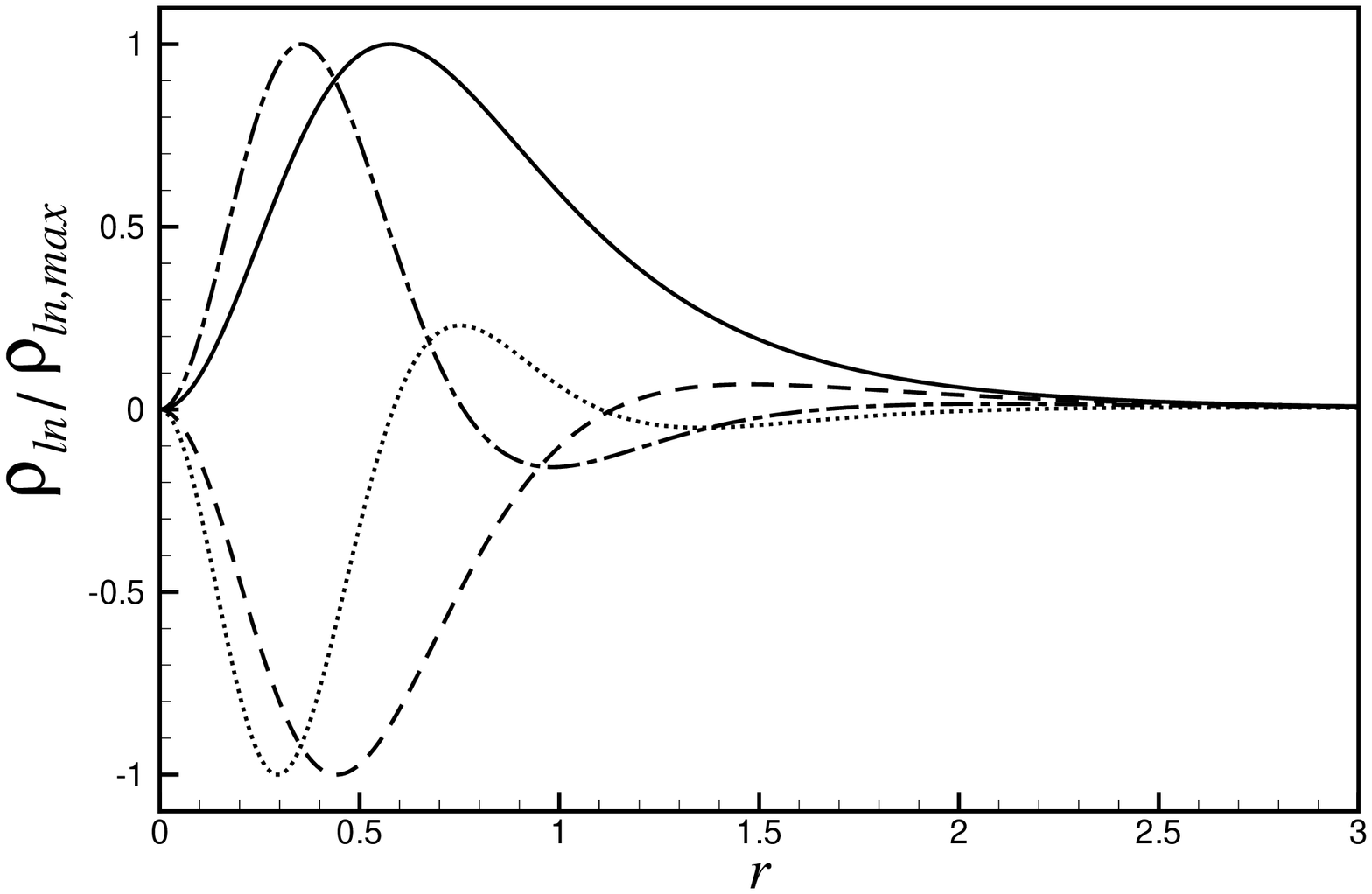}
  \end{tabular}
\caption{Potential (left) and density (right) basis functions for 
$l=2$. Solid, dashed, dash-dotted and dotted lines correspond to 
$n=$0, 1, 2, and 3, respectively. Densities have been scaled to 
their maximum values.}
 \label{curve}
\end{figure}
We have also examined the performance of our basis set in modeling 
the spherical and oblate Kuzmin-Kutuzov models. Our experiments show 
that taking a handful of expansion terms reconstructs the original 
model with a relative error less than \%1.

\vspace{-0.15in}

\section{Conclusions}\label{sec:conclusion}
Our basis functions constitute a bi-orthogonal set over the domain 
$0\le r < \infty$. By adjusting the length scale $a$, it is possible to 
model stellar systems of different core radii and spatial extensions. 
The wave length of our radial basis functions decreases towards the 
center while their wave amplitudes are not changed substantially. 
These properties are useful for the instability analysis 
of three dimensional stellar systems, for instabilities occur in central 
regions of stellar systems and accurate calculation of perturbed 
physical quantities demands high-resolution functions near the center.

\end{document}